\documentclass[pra,twocolumn,titlepage,nofootinbib,amsmath%
,preprintnumbers%
]{revtex4}%
\usepackage{amsmath}
\usepackage{amsfonts}
\usepackage{amssymb}
\usepackage{graphicx}%
\usepackage{color}

\begin{document}

\title{The Lamb shift of the $1s$ state in hydrogen: two-loop and three-loop contributions}
\author{Savely~G.~Karshenboim}
\email{savely.karshenboim@mpq.mpg.de}
\affiliation{Ludwig-Maximilians-Universit{\"a}t, Fakult{\"a}t f\"ur Physik, 80799 M\"unchen, Germany}
\affiliation{Max-Planck-Institut f\"ur Quantenoptik, Garching, 85748, Germany}
\affiliation{Pulkovo Observatory, St.Petersburg, 196140, Russia}
\author{Akira Ozawa}
\affiliation{Max-Planck-Institut f\"ur Quantenoptik, Garching, 85748, Germany}
\author{Valery A. Shelyuto}
\affiliation{D.~I. Mendeleev Institute for Metrology, St.Petersburg,
190005, Russia}
\affiliation{Pulkovo Observatory, St.Petersburg, 196140, Russia}
\author{Robert Szafron}
\affiliation{Technische Universit\"at M\"unchen, Fakult\"at f\"ur Physik, 85748 Garching, Germany}
\author{Vladimir G. Ivanov}
\affiliation{Pulkovo Observatory, St.Petersburg, 196140, Russia}


%

\preprint{TUM-HEP-1179/18}

\begin{abstract}
We consider the $1s$ Lamb shift in hydrogen and helium ions, a quantity, required for an accurate determination of the Rydberg constant and the proton charge radius by means of hydrogen spectroscopy, as well as for precision tests of the bound-state QED. The dominant QED contribution to the uncertainty originates from $\alpha^8m$ external-field contributions (i.e., the contributions at the non-recoil limit). We discuss the two- and three-loop cases and in particular, we revisit calculations of the coefficients $B_{61}, B_{60}, C_{50}$ in standard notation.

We have found a missing logarithmic contribution of order $\alpha^2(Z\alpha)^6m$. We have also obtained leading pure self-energy logarithmic contributions of order $\alpha^2(Z\alpha)^8m$ and $\alpha^2(Z\alpha)^9m$ and estimated the subleading terms of order $\alpha^2(Z\alpha)^7m$, $\alpha^2(Z\alpha)^8m$, and $\alpha^2(Z\alpha)^9m$. The determination of those higher-order contributions enabled us to improve the overall accuracy of the evaluation of the two-loop self-energy of the electron.

We investigated the asymptotic behavior of the integrand related to the next-to-leading three-loop term (order $\alpha^3(Z\alpha)^5m$, coefficient $C_{50}$ in standard notation) and applied it to approximate integration over the loop momentum. Our result for contributions to the $1s$ Lamb shift for the total three loop next-to-leading term is $(-3.3\pm10.5)(\alpha^3/\pi^3)(Z\alpha)^5m$.

Altogether, we have completed the evaluation of the logarithmic contributions to the $1s$ Lamb shift of order $\alpha^8m$ and reduced the overall $\alpha^8m$ uncertainty by approximately a factor of three for H, D, and He$^+$ as compared with the most recent CODATA compilation.
\end{abstract}

\maketitle

\section{Introduction\label{s:int}}

Already for a few years, there exists a discrepancy in the determination of the proton charge radius  by means of the spectroscopy of ordinary and muonic hydrogen (see, e.g., \cite{muh1,codata2014}), commonly known as the proton radius puzzle.
There are different contributions to the uncertainty of the determination of the proton radius by those methods.
The largest uncertainty originates from the hydrogen spectroscopy and a serious experimental activity in this direction is in progress (see, e.g., \cite{hessel:lp,lkb1s3s,mpq2s4p}). The second largest uncertainty comes from the Quantum electrodynamics (QED) theory of the $1s$ Lamb shift in hydrogen \cite{codata2014}. There are a few theoretical problems which require clarification. They relate to two-loop and three-loop radiative corrections. Some higher-order contributions have not been cross-checked, and some not studied at all. In particular, the two-loop contributions of order $\alpha^2(Z\alpha)^5m$ \cite{b50se,LbL1} are well established, while at the next order in $Z\alpha$ the contributions for the virtual light-by-light scattering have not been studied properly (see, e.g., a discussion on a previously missed term in \cite{LbL:CS}). Meanwhile, the results for the pure self-energy contribution of order $\alpha^2(Z\alpha)^6m$ \cite{jentscura1s,yerokhin09} are to some extent controversial (see, e.g., \cite{codata2014}). One more challenge is related to the next-to-leading order three-loop contribution (order $\alpha^3(Z\alpha)^5m$); the existing estimation \cite{codata2014} does not have a solid ground.

Besides the proton radius puzzle, an improvement of the theoretical prediction of the $1s$ Lamb shift is essential for the determination of the Rydberg constant \cite{muh1,codata2014}, precision tests of the bound-state QED, constraints on light neutral particles, such as a dark photon from physics of simple atoms (see, e.g., \cite{constraint}), and interpretation of the currently ongoing $1s-2s$ He$^+$ experiments~\cite{he_ion1,he_ion2}.

The Lamb shift of the atomic energy levels is a QED effect, that can be experimentally studied in light hydrogen-like atoms with 
high accuracy (cf. \cite{mpq_h_new_1}). The theoretical prediction of this phenomenon involves the values of the input parameters, such as the Rydberg constant and the proton charge radius, that limit the accuracy of the calculations. A separate input to the uncertainty originates in the computation of various high-order QED effects. The dominant contributions to the QED error budget come from the radiative
corrections in the external-field approximation. We follow the standard convention and parametrize these corrections as (see, e.g., \cite{codata2014,VASH-book})
\begin{equation}\label{eq:F123}
\Delta E(ns)=\frac{\alpha \,(Z\alpha)^4m}{\pi \, n^3}
\left(F^{(1)}+\frac{\alpha}{\pi} \, F^{(2)}+\left(\frac{\alpha}{\pi}\right)^2F^{(3)}+\dots\!\right)\!,
\end{equation}
where $F^{(i)}=F^{(i)}_{ns}(Z\alpha)$ corresponds to the $i$-loop radiative insertions and the relevant contributions are at the one-, two-, and three-loop level. The four-loop contributions are neglected in (\ref{eq:F123}). The uncertainty due to the unknown leading four-loop term, which is expected at the level of a few units of $\alpha^4/\pi^4(Z\alpha)^4m$, is essentially below the uncertainty of the higher-order two-loop and three-loop terms. The latter are at the level of ten units of $\alpha^2/\pi^2(Z\alpha)^6m$ and $\alpha^3/\pi^3(Z\alpha)^5m$, respectively (see below).

Theory of the one-loop contributions is firmly established (see \cite{codata2014,VASH-book} for details). The largest and most important contribution, related to the electron self-energy, has been calculated directly for $Z=1,2$ \cite{1sse1}, i.e., for H and He$^+$. We consider below the two-loop and three-loop radiative corrections.

The functions $F^{(i)}$ can be expanded at low $Z\alpha$ and at two and three loops, the results read
\begin{eqnarray}
(Z\alpha)^4\,F^{(2)}(nl)&=&\sum_{kp}B_{kp}(Z\alpha)^k\ln^p{\frac{1}{(Z\alpha)^2}} \;, \label{twoloop}\\
(Z\alpha)^4\,F^{(3)}(nl)&=&\sum_{kp}C_{kp}(Z\alpha)^k\ln^p{\frac{1}{(Z\alpha)^2}}\;. \label{threeloop}
\end{eqnarray}
Here, we focus on the~$1s$ state and the $F^{(i)}$ coefficients are always meant to be related to the aforementioned, $1s$ state.

It is not clear {\em a priori\/} which logarithmic terms are present in (\ref{twoloop}) and (\ref{threeloop}).  Sometimes a special study is required.
For example, it was believed \cite{codata2014} until recently that $C_{63}\neq0$, while the presence of $B_{72}\neq0$ was rather disputable. Both issues have been recently resolved in \cite{our_b72} and we discuss it also below.

A number of the two-loop ($B_{\dots}$) and three-loop ($C_{\dots}$) coefficients have been known with a sufficient accuracy. These include $B_{40}$, $B_{50}$, $B_{63}$, $B_{62}$, $B_{61}$, and $C_{40}$. Estimations with a credible uncertainty have also been available for $B_{60}, C_{50}$, and $C_{63}$. A concise summary concerning all these coefficients can be found in \cite{codata2014}. Some of the corrections have been revisited since publication of \cite{codata2014}. These include, e.g., $B_{61}$ \cite{LbL:CS} and $B_{72}, C_{63}$, and $C_{62}$ \cite{our_b72}.

In this letter, we reconsider $B_{61}$ (order $\alpha^2(Z\alpha)^6m\ln(Z\alpha)$), $B_{60}$ (the non-logarithmic $\alpha^2(Z\alpha)^6m$ term), and $ C_{50}$ ($\alpha^3(Z\alpha)^5m$) and discuss them below  in subsequent sections in detail.
Our findings are summarized in Tables~\ref{t:sum:two} and \ref{t:sum:three}.

\begin{table}[htbp]
\begin{center}
\begin{tabular}{l|c|c|c}
\hline
Quantity & $B_{61}^{\rm tot}$ &  $B_{60}^{\rm tot}$ & $G_{60}^{\rm tot}(Z=1)$  \\
\hline
\cite{codata2014}: coefficient  & $48.958\,590$ & $-81.3(19.7)$ & \\
contribution, kHz & 48.50 & $-8.2(2.0)$ & \\
\hline
this work: coefficient  & $49.788\,899$ & & $-94.5(6.6)$ \\
contribution, kHz  & 49.32 & & $-9.5(0.7)$ \\
\hline
\end{tabular}
\caption{Two-loop coefficients and their contributions to the~$1s$ Lamb shift in hydrogen. $G_{60}^{\rm tot}(Z)$ is equal to $B_{60}$ together with all the higher-order (in $Z\alpha$) corrections (see (\ref{G:twoloop})).
\label{t:sum:two}}
\vspace{-6.0mm}
\end{center}
\end{table}

In the case of the two-loop corrections, rather than $B_{60}$, we use $G_{60}(Z\alpha)$ defined as
\begin{equation}
G_{60}(Z\alpha)=B_{60}+\sum_{kp; k\geq7}B_{kp}(Z\alpha)^{k-6}\ln^p{\frac{1}{(Z\alpha)^2}} \;, \label{G:twoloop}
\end{equation}
i.e., it is equal to $B_{60}$ with all the higher-order (in $Z\alpha$) corrections included. $G_{60}(Z\alpha)$ is more appropriate if one uses results of numerical calculations.

\begin{table}[htbp]
\begin{center}
\begin{tabular}{l|c|c|c}
\hline
Quantity & $C_{50}^{\rm tot}$ &  $C_{63}^{\rm tot}$ & $C_{62}^{\rm tot}$  \\
\hline
\cite{codata2014}: coefficient  & $\pm30$ & $\pm1$ & \\
\cite{codata2014}: contribution, kHz & $\pm 0.96$ & $\pm0.22$ & \\
\hline
this work$^*$: coefficient & $-3.3(10.5)$ & 0 & $-0.36$\\
this work$^*$: contribution, kHz  & $-0.11(34)$ & 0 & $-0.01$\\
\hline
\end{tabular}
\caption{Three-loop coefficients and their contributions to the $1s$ Lamb shift in hydrogen. $^*$We use here $C_{63,62}$ from~\cite{our_b72}.
\label{t:sum:three}}
\end{center}
\vspace{-6.0mm}
\end{table}


\section{Additional logarithmic two-loop contributions in order $\alpha^2(Z\alpha)^6m$ \label{s:log}}

\begin{figure}[thbp]
\begin{center}
\resizebox{0.80\columnwidth}{!}{\includegraphics[clip]{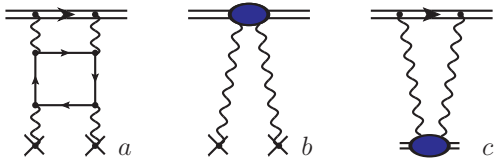}}
\end{center}
\vspace{-3.0mm}
\caption{Example diagrams for the LbL contributions: an `initial' diagram ($a$) and two effective diagrams ($b,c$). The double horizontal line is for the Coulomb propagator,
The effective diagrams are result of the hard integrations (with momenta comparable with $m$), which produce effective point-like vertices, while the remaining photons are soft (i.e. the momenta are much smaller than $m$).
}
\label{f:lbl}       
\vspace{-4.0mm}
\end{figure}

We begin with the two-loop logarithmic coefficient $B_{61}$. First calculated in~\cite{b601s,log1}, the result was applied in~\cite{codata2014}. After original publication, the diagrams with the light-by-light (LbL) scattering block (see Fig.~\ref{f:lbl}$a$) have been studied, and a correction to the previous result was found~\cite{LbL:CS} due to the LbL diagrams  overlooked in \cite{b601s,log1}. The LbL contributions are the most difficult for the numerical calculations. The analytic calculations have been available only for the order $\alpha^2(Z\alpha)^5m$ \cite{LbL1} and absent for a while for the next order in $Z\alpha$ until the publication of \cite{LbL:CS}. Those diagrams receive a contribution from soft photons responsible for the appearance of a long-distance potential.
After integrating out the hard modes (i.e. with momenta comparable with $m$), effective local operators appear, which give rise to the two-photon vertices shown in diagrams b and c in Fig.~\ref{f:lbl}. The remaining photons are soft (i.e. their momenta are much smaller than $m$).
There are two possible soft pairs of photons, those which connect the nucleus and the electron loop (see Fig.~\ref{f:lbl}$b$) and those which connect 
the electron line and the electron loop (see Fig.~\ref{f:lbl}$c$).
The former case was covered by \cite{LbL:CS}, while we consider the latter here.

The bottom part of the diagram in Fig.~\ref{f:lbl}$a$, i.e., the electron loop in the Coulomb field of the nucleus, is the known virtual Delbr\"uck scattering amplitude (see, e.g., \cite{rev:vD2,vDs} and references therein). The upper part, i.e., the electron line and two soft photons connecting the electron line and the electron loop, can be
drastically simplified within the soft-photon kinematics, where the energy transfer ($q_0$) is comparable with the momentum transfer ($\bf q$)  and $Z\alpha m \ll |q_0| \sim |{\bf q}|\ll m$. The integral over $q_0$ simplifies considerably within a kind of static-electron kinematics discussed in details in
\cite{LbL1mu}.

The resultant integral induces an effective potential that behaves as $r^{-4}$ (cf. \cite{our_mulbl}, see also \cite{LbL:CS}). In the $1s$ state, the expectation value of this potential diverges at
the short end of the interval $1/m\ll r \ll 1/(Z\alpha m)$.
The manifestation of the divergence in the perturbation theory is a logarithmically enhanced correction to the hydrogen energy levels. On the technical side, the calculation is similar to that in \cite{LbL:CS} if we use the effective field theory approach. To confirm our result, we also considered diagrams with triple photon exchange and extracted the logarithmically divergent part; both methods gave the same correction to the $B_{61}$ coefficient.
The total result for the logarithmic LbL contribution, including the one from~\cite{LbL:CS}, reads
\begin{equation}\label{eq:lbl:log}
\Delta E^{\rm LbL}(1s) =\frac{\alpha^2(Z\alpha)^6m}{\pi^2}\left[
\frac{709\pi^2}{3456}-\frac{43}{36}
\right]\ln\frac1{(Z\alpha)^2}
\,.
\end{equation}
Result (\ref{eq:lbl:log}) is over 2.5 times larger than that of the previous [partial] computation~\cite{LbL:CS}. In standard notation (cf. \cite{codata2014}) it corresponds to the $B_{61}$ coefficient.

\section{Two-loop contributions with closed electron loops in order $\alpha^2(Z\alpha)^6m$ \label{s:closed}}

\begin{figure}[thbp]
\begin{center}
\resizebox{0.8\columnwidth}{!}{\includegraphics[clip]{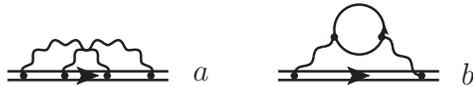}}
\end{center}
\vspace{-2.0mm}
\caption{Example diagrams for the two-loop contributions to the Lamb shift:
a pure self-energy one ($a$) and
and one
with an electron loop in the free-loop approximation ($b$).}
\label{f:two}       
\vspace{-4.0mm}
\end{figure}

While considering the non-logarithmic part of the $\alpha^2(Z\alpha)^6m$ correction, i.e., the coefficient $B_{60}$ and higher-order terms, one has to distinguish three groups of diagrams and treat them differently. One group originates from the `pure' self-energy (SE) diagrams, i.e., the diagrams without any closed electron loops (see Fig.~\ref{f:two}$a$). The remaining groups, on the other hand, include the closed electron loops. The second group contains the loops in the so-called free-loop approximation, i.e., all appearing closed electron loops are due to the vacuum polarization (see Fig.~\ref{f:two}$b$). The last group contains virtual LbL scattering subdiagrams (see, e.g., Fig.~\ref{f:lbl}$a$). (In high-$Z$ atomic physics, those diagrams are referred to as the vacuum polarization in the presence of the Coulomb field of a nucleus.)

The most accurately computed results exist for the {\em free\/}-loop approximation diagrams studied in~\cite{yero_num_1}. The result reads
\begin{eqnarray}
G_{60}^{\rm free}(Z\!=\!1)&=&-15.0(4)\;,\nonumber\\
G^{\rm free}_{60}(Z\!=\!2)&=&-13.9(1)
\;.\end{eqnarray}

The contributions beyond the free-loop approximation at order $\alpha^2(Z\alpha)^6m$ belong to two groups. One is due to radiative corrections to the Wichmann-Kroll contribution. (The Wichmann-Kroll contribution by itself is of order $\alpha(Z\alpha)^6m$.) We estimate it as
\begin{equation}\label{eq:b60:rwk}
B_{60}^{\rm rWK}(ns)=0.13\pm0.13\;.
\end{equation}
The estimation is based on a similarity of the behavior of a radiative correction to the Wichmann-Kroll potential and the K\"allen-Sabry potential in the so-called $t$ channel.

The other group arises due to  Coulomb corrections to the LbL contribution of order $\alpha^2(Z\alpha)^5m$. We have already considered their logarithmic part above in (\ref{eq:lbl:log}). We estimate the non-logarithmic part as
\begin{equation}\label{eq:b60:lbl}
B_{60}^{\rm LbL}=\pm \pi B_{61}^{\rm LbL}\simeq \pm 2.6\;.
\end{equation}
The $B_{60}$ term beyond the free-loop approximation was previously estimated in \cite{yero_num_1}. However, it was based on incorrect assumptions about the logarithmic contributions for the diagrams beyond the free-loop approximation, and thus we do not take into consideration those estimates. The quantitatively largest contribution in our consideration of the $B_{60}$ term beyond the free-loop approximation comes as a `tail' of the logarithmic $B_{61}$ term. The summary of the individual contributions to $G_{60}(1s)$ is given in Table~\ref{t:b:all2loop}.

\begin{table}[htbp]
\begin{center}
\begin{tabular}{l|c|c|c||c}
\hline
Quantity  &$G_{60}^{\rm SE}(1s)$ & $G_{60}^{\rm free}(1s)$ & $G_{60}^{\rm beyond}(1s)$ & $G_{60}^{\rm tot}(1s)$\\
\hline
$Z=1$ & $-79.6(6.0)$ & $-15.0(4)$ & 0.1(2.6)& $-94.5(6.6)$ \\
$Z=2$ & $-83.3(5.2)$ & $-13.9(1)$ & 0.1(2.6)& $-97.1(5.8)$ \\
\hline
\end{tabular}
\caption{Higher-order two-loop contributions to the $1s$ Lamb shift in hydrogen and the helium ion. The {\em free\/}-loop approximation result is from \cite{yero_num_1}. The pure {\em SE\/} value as well as the contribution {\em beyond\/} the free-loop approximation is a result of this letter.
\label{t:b:all2loop}}
\end{center}
\vspace{-6.0mm}
\end{table}

The estimate above is obtained by a suggestion that a natural magnitude of the constant  accompanying a logarithm is $\pi$, which is inspired by the value of the imaginary part of the logarithm of a negative real number. In a term with several logarithms, we substitute each of them by $\pi$, which produces a combinatoric factor. Often the terms beyond the leading logarithmic term are estimated by 50\% of its value. Using $\pi$ and combinatoric factors, we estimate the subleading terms in the case of the leading single-logarithm as below 50\%, but in the case of leading double- or triple-logarithmic term as above 50\%. We think that it is more realistic than a naive 50\%-estimate for all the separate cases.

\section{Pure self-energy two-loop contributions $\alpha^2(Z\alpha)^6m$ \label{s:se}}

The situation concerning the pure SE part of $B_{60}$ is more complicated than that of the closed-electron-loop contributions. A partial calculation exists, and it is accompanied by a plausible estimate of the unknown part of the contribution~\cite{jentscura1s},
\begin{equation}\label{eq:b60:j}
B_{60}^{\rm pure\;SE}=-61.6(9.2)\;.
\end{equation}
The large magnitude of the $B_{60}^{\rm pure\;SE}$ coefficient is due to an enhancement of the low-momentum contribution, while the uncertainty comes from the unknown high-momentum one.
Suggesting that the missing high-momentum contribution is not enhanced, we arrive at the result of $\pm\pi^3B_{63}$ for the missing contribution that coincides with the uncertainty in (\ref{eq:b60:j}). Consequently, our estimation of the magnitude of the unknown terms in~(\ref{eq:b60:lbl}) is consistent with that in~\cite{jentscura1s}.

There exist essentially three approaches to calculation of the higher-order two-loop contributions. One suggests the use of an $Z\alpha$ expansion, in which case the accuracy is limited by (\ref{eq:b60:j}) \cite{jentscura1s}. It is also necessary to know the higher-order logarithmic terms, such as \cite{our_b72}
 \begin{equation}\label{eq:b72}
B_{72}^{\rm pure\;SE}=-\frac{2\pi}{3}\,\left(\frac{139}{32}-2\ln2\right)\;.
\end{equation}
The size of the logarithmic contribution is smaller than of the uncertainty above, but not negligible.

The second approach uses exact in $Z\alpha$ numerical calculations at $Z=1,2$. In the case of two-loop contributions, that approach has been successfully applied for the contributions with closed electron loops in the free-loop approximation \cite{yero_num_1} (see above), but its application to pure SE diagrams has proved challenging. Only the results at medium $Z$ such as $Z=10,12,15,17,20,25,30$ \cite{yerokhin09} 
are available. Those still can be extrapolated to $Z=0,1,2$.
(The third approach includes a fit using the result of (\ref{eq:b72}) as a data point at $Z=0$ for $G_{60}^{\rm SE}(Z)$ from (\ref{G:twoloop}).)

Due to the low accuracy such an extrapolation is possible for $F^{(2)}(Z\alpha)$, only because a number of the leading coefficients, such as ($B_{40}, B_{50}, B_{63}, B_{62}$, and $B_{61}$) is known (see \cite{codata2014,VASH-book} and references therein). In the meantime the `data area' ($Z=10,12,15,17,20,25,30$) is relatively far from the `target area' ($Z=0,1,2$) and it contains relatively few data points. The logarithmic terms go through a bigger change on their way from the data area to the target area than within the data area. Accordingly, from the point of view of a phenomenological fit, we have to consider nearly coinciding fits with different extrapolation expectations. Since we need not  only  to fit the data but eventually also to extrapolate, we have to maintain the correct shape of the fit function (see (\ref{twoloop})).

To deal with logarithmic terms at orders $\alpha^2(Z\alpha)^7m$ and $\alpha^2(Z\alpha)^8m$, a calculation of some and an estimation of others is necessary. We present the summary of the leading logarithmic terms at each order in $Z\alpha$ in Table~\ref{t:b:coef:0}. The coefficients $B_{84}$ and $B_{93}$ are calculated in this letter using techniques developed in~\cite{log3,log2vp,our_b72}.

\begin{table}[htbp]
\begin{center}
\begin{tabular}{l|c|c|c|c
}
\hline
Coefficient  &$B_{63}$ & $B_{72}$ & $B_{84}$ & $B_{93}$ 
\\
\hline
Value  & $-8/27$ & $-6.19$ & $-7/27$  & $5/6\cdot B_{72}\simeq-5.162$ 
\\
\hline
\end{tabular}
\caption{The leading higher-order pure SE two-loop logarithmic contributions. Note: the leading logarithmic terms of orders $\alpha^2(Z\alpha)^6m$ \cite{log3} and $\alpha^2(Z\alpha)^8m$
come only from the pure self-energy. In contrast to that, the leading logarithms of order $\alpha^2(Z\alpha)^7m$ \cite{our_b72} and $\alpha^2(Z\alpha)^9m$
come both from the diagrams with and without closed electron loops.
Here we present only their self-energy part.
\label{t:b:coef:0}}
\end{center}
\vspace{-6.0mm}
\end{table}

To estimate the subleading terms we use several approaches. As for an `estimation' we understand a constraint with a relatively large uncertainty such as in~(\ref{eq:b60:j}),
which allows us to obtain more than one estimate for each subleading coefficient. Ultimately, we choose the most conservative constraint for the coefficients.
The summary of our estimations is given in Table~\ref{t:b:coef:2}.

The coefficients can be used for both for a low-$Z\alpha$ expansion and fits. We have to explain how we used the constraints for the fits. We consider the constraints as additional data and include their deviation from the related central values, measured in the units of their uncertainties, into the final $\chi^2$ we have to minimize. That is, e.g., similar to a treatment applied in \cite{codata2014} for the various not very precise theoretical corrections. Such a fitting procedure allows one to easily combine theoretical constraints with the existing `true' data.

\begin{table}[htbp]
\begin{center}
\begin{tabular}{l|c|c|c|c|c|c}
\hline
Coefficient & $B_{71}$  & $B_{70}$ &$B_{83}$&$B_{82}$&$B_{81}$&$B_{80}$ \\
\hline
Value  &   $-12(40)$ & $\pm72$ &  $\pm 3.2$ &$\pm 50$&$\pm 150$ &$\pm 200$\\
\hline
\end{tabular}
\caption{Estimated values of the coefficients for two-loop pure SE subleading terms in order $\alpha^2(Z\alpha)^7m$ and $\alpha^2(Z\alpha)^8m$.
\label{t:b:coef:2}}
\end{center}
\vspace{-6.0mm}
\end{table}

Often, for the higher-order terms, it is possible to estimate the magnitude plausibly, but not the sign of a coefficient and therefore frequently the central values of estimations are zero. Once $B_{71,70}$ are estimated, we can find the result of the low-$Z\alpha$ expansion of $G_{60}^{\rm SE}(Z\!=\!1,2)$ (see Table~\ref{t:b:res}).

\begin{table}[htbp]
\begin{center}
\begin{tabular}{l|c|c|c}
\hline
Quantity  &$B_{60}^{\rm SE}$ & $G_{60}^{\rm SE}(Z=1)$ & $G_{60}^{\rm SE}(Z=2)$\\
\hline
Low-$Z\alpha$ expansion  & $-61.6(9.2)$ & $-66.8(9.6)$ & $-69.6(10.5)$ \\
Fit over data from \cite{yerokhin09}  & $-90(12)$ & $-94(10)$ & $-95(9)$ \\
Combined fit  & $-72.4(7.2)$ & $-79.6(6.0)$ & $-83.3(5.2)$ \\
\hline
\end{tabular}
\caption{Higher-order two-loop pure self-energy contribution to the $1s$ Lamb shift in hydrogen and the helium ion. The {\em combined\/} fit includes the numerical data from \cite{yerokhin09} and the value of $B_{60}$ from \cite{jentscura1s}, and less accurate numerical results from \cite{yero_num_3}.
\label{t:b:res}}
\vspace{-6.0mm}
\end{center}
\end{table}

To compare those low-$Z\alpha$ results with the numerical data, we need to fit them. We found that by setting $B_{72}=0$, the fit results for $B_{60}$ are  shifted by $8-10$ from the value $B_{60}$ of (\ref{eq:b60:j}). That would put the fit and that value of $B_{60}$ in disagreement and would not allow a combined fit. All the previously used fits have ignored the double-logarithmic $B_{72}$ term. Consequently, the fits found in the literature use an unrealistic shape with no estimation of systematic effects (see, e.g., \cite{yerokhin09}). Consequently, a comparison with the previously performed fits is meaningless. We have performed a fitting of the numerical data \cite{yerokhin09} ourselves, using realistic approximation functions. We present the result in Table~\ref{t:b:res} including the results of the combined fit, i.e., a fit which includes the low-$Z\alpha$ constraints and numerical data from \cite{yerokhin09}. We consider a difference between the low-$Z$ value and the fit over the numerical data, which is somewhat below $2\,\sigma$ as a fair agreement which validates the use of a combined fit.

The summary for the calculation of the two-loop contributions in the external field approximation is given in Table~\ref{t:sum:two}.

\section{Next-to-leading three-loop contributions \label{s:three}}

The three-loop theory is more complicated and less advanced than the two-loop one. Only its leading term to the Lamb shift (order $\alpha^3(Z\alpha)^4m$, coefficient $C_{40}$) is known \cite{c40_dirac,c40_pauli,c40_vp}. The next-to-leading one (order $\alpha^3(Z\alpha)^5m$, $C_{50}$) has been calculated only partially \cite{some_c50} and boldly estimated in \cite{codata2014}. After improvement of the accuracy of $B_{60}$ above, $C_{50}$ \cite{codata2014} becomes the largest source of the QED uncertainty for the $1s$ Lamb shift in hydrogen.

\begin{figure}[thbp]
\begin{center}
\resizebox{0.30\columnwidth}{!}{\includegraphics[clip]{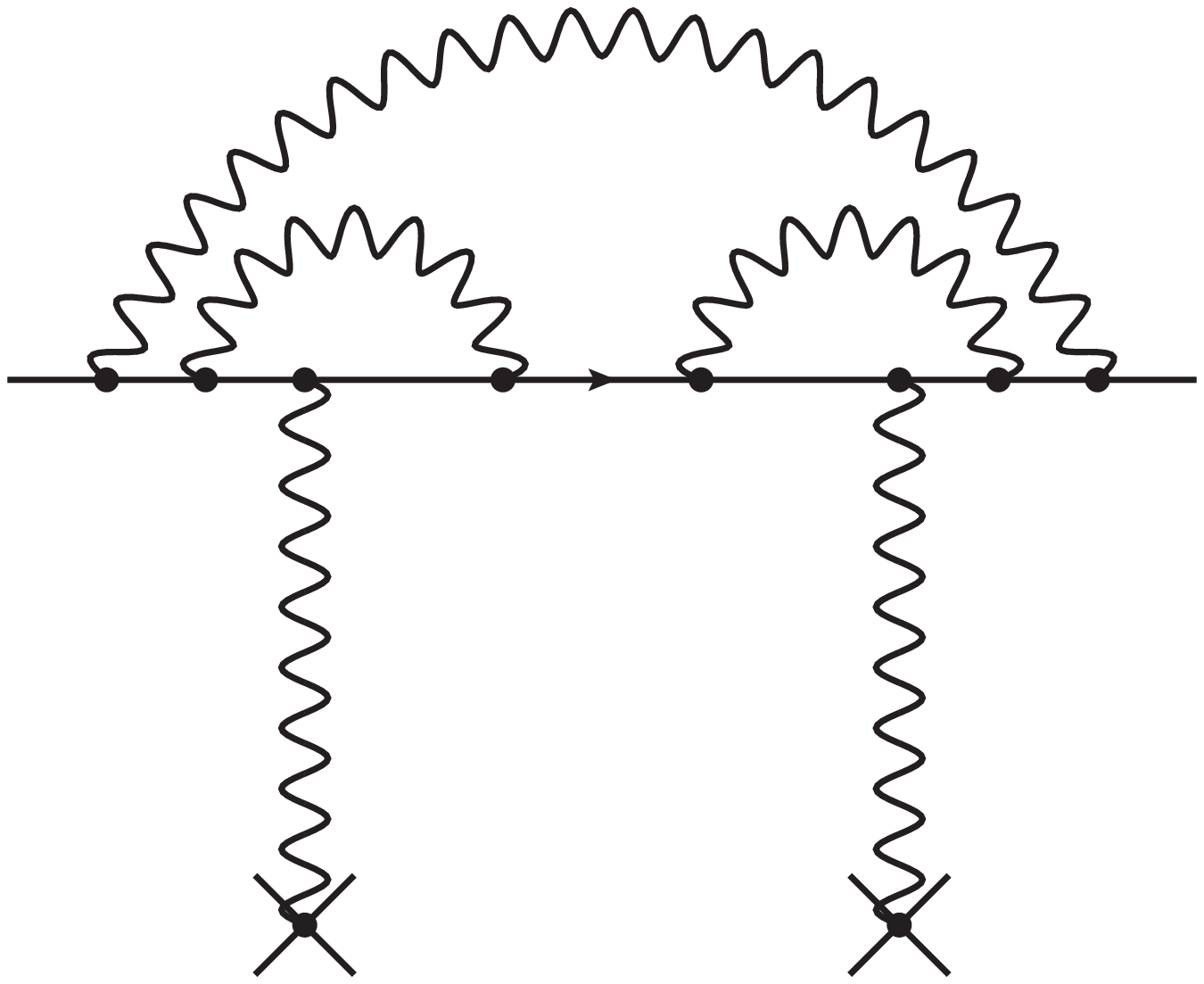}}
\end{center}
\vspace{-2.0mm}
\caption{An example diagram for the three-loop contribution to the Lamb shift.}
\label{f:three}       
\vspace{-4.0mm}
\end{figure}

The $\alpha^3(Z\alpha)^5m$ contribution can be represented as a set of two-photon exchange diagrams (see Fig.~\ref{f:three}). The related expression may be written in terms of the integral over the loop momentum $q$ (cf. \cite{log3})
\begin{equation}\label{eq:c50:t}
\int_0^\infty{\frac{dq}{q^4}}\,T(q^2)
\;,
\end{equation}
where $T(q^2)$ is a radiative correction to the skeleton-diagram integral, which is related to a virtual forward Compton scattering  amplitude.

The calculation of the radiative factor $T(q^2)$ is very complicated. Here, we calculate its asymptotics at high and low $q$ and estimate the total integral, by integrating those asymptotic expressions. As mentioned previously,  a part of the contributions, i.e., the diagrams with closed electron loops in free-loop approximation except for graphs with the two-loop pure self-energy with one electron vacuum-polarization insertion, have already been considered in \cite{some_c50}. Here we estimate the unknown diagrams by integrating the asymptotics of the related integrand.
The complete three-loop result is
\begin{equation}\label{eq:c50:tot}
C_{50}^{\rm total}(ns)=-3.3(10.5)\;.
\end{equation}

To verify our method, we have also found 
the contributions of order $\alpha(Z\alpha)^5m$ and $\alpha^2(Z\alpha)^5m$. Our estimation is in a perfect agreement with known results \cite{a50,b50se}. In each case of interest (one loop, two loops, three loops) the asymptotics of $T(q^2)$ are of the same sign for high and low $q$. That is an important requirement for a reliable estimation of the integral through the asymptotics of the integrand.

The present situation with the three-loop contributions is summarized in Table~\ref{t:sum:three}. The $C_{50}$ uncertainty is reduced by a factor of 3. This makes the $C_{50}$ uncertainty comparable with the CODATA's $C_{63}$ one in \cite{codata2014}. Fortunately, the latter was eliminated in \cite{our_b72}, where it was found that
\begin{eqnarray}
C_{63}&=&0\;,\nonumber\\
C_{62}&\simeq& -0.36\;.
\end{eqnarray}

\section{Summary and conclusions \label{s:summary}}

The summary on the theoretical accuracy of the $1s$ Lamb shift calculation for light hydrogen-like atoms with $Z=1,2$ is presented in Table~\ref{t:sum:h:he}.
The uncertainty from the external-filed contributions, considered in this paper, is due to $\alpha^8m$ terms and consists of two sources, one is two-loop's $G_{60}$ and the other is three-loop's $C_{50}$. A comparison with the existing calculations of other authors is given in the introduction, in Tables~\ref{t:sum:two} and \ref{t:sum:three} in terms of the related coefficients and absolute values of the contributions for hydrogen. As one can see from there both two-loop and three-loop uncertainties are reduced approximately by factor of three.

\begin{table}[htbp]
\begin{center}
\begin{tabular}{l|c|c|c}
\hline
Contribution, kHz  & $G_{60}^{\rm tot}$ &  $C_{50}^{\rm tot}$ & RR16  \\
\hline
Contribution for H  & $-9.5(0.7)$  & $-0.11(34)$ & $1.5(1.0)$ \\
Contribution for D  & $-9.5(0.7)$  & $-0.11(34)$ & $0.76(0.49)$\\
Contribution for ${}^3$He$^+$  & $-625(37)$ & $-3.4(10.8)$ & $23(18)$\\
Contribution for ${}^4$He$^+$  & $-625(37)$ & $-3.4(10.8)$ & $18(14)$\\
\hline
\end{tabular}
\caption{The most uncertain contributions to the $1s$ Lamb shift in hydrogen, deuterium and the helium ions. RR16 stands for the $\alpha(Z\alpha)^6m^2/M$ radiative-recoil contribution, which is known only in the leading logarithmic approximation (see (\ref{eq:rr})).
\label{t:sum:h:he}}
\end{center}
\vspace{-6.0mm}
\end{table}

The dominant contribution to the uncertainty budget for hydrogen currently comes from the radiative-recoil contribution of order $\alpha(Z\alpha)^6m^2/M$, that is known in the leading logarithmic approximation \cite{aza6ln22,aza6ln21}
\begin{equation}\label{eq:rr}
\Delta E_{\rm RR16}(1s)=\frac23 \frac{\alpha(Z\alpha)^6m}{\pi}\,\frac{m}M\,\ln^2{\frac1{(Z\alpha)^2}}\;.
\end{equation}
The uncertainty in Table~\ref{t:sum:h:he} comes from an estimation of subleading terms.
For its estimation we use here the approach with $\pi$'s and combinatoric coefficients, as explained above, and the uncertainty is somewhat above 50\% (cf. \cite{codata2014}).
The key uncertainty sources in \cite{codata2014} have also included pure recoil corrections, but their uncertainty (of about 0.7\;kHz for H) has recently been eliminated \cite{recoilyero} by a direct calculation of the recoil corrections for $Z=1,2$.

Concluding, we have revisited the theory of the $\alpha^8m$ contributions to the Lamb shift of the $1s$ state in hydrogen and deuterium atoms and helium ions. We completed the calculation of the logarithmic terms, considered a controversy in the non-logarithmic two-loop contribution and improved its accuracy by approximately a factor of three, and obtained a complete approximate result for the three-loop terms, which is more reliable and three times more accurate than a previous bold estimation.

The most accurate experimental results are available for the $1s-2s$ transition in hydrogen and deuterium~\cite{mpq_h_new_1,
mpq_d_new}. Experimental efforts to measure the $1s-2s$ transition in the helium ion are underway \cite{he_ion1,he_ion2}. Since the weight of the individual contributions to the theoretical uncertainty budget varies substantially (see Table~\ref{t:sum:h:he}), combining the hydrogen and helium-ion experimental results would be beneficial not only for hydrogen and helium-ion spectroscopy but also for various applications including precision tests of bound-state QED, determination of the Rydberg constant, and constraints on new light neutral particles.

A complete and detailed derivation, covering the technical side of the computations of our new results presented in this letter, is under preparation and will be published elsewhere.

The authors are grateful to A. Czarnecki, M.I. Eides, K. Eikema, V.I. Korobov,
E.Yu. Kor\-zinin, K. Pachucki, Th. Udem, and V.A. Yerokhin for valuable stimulating discussions.
The work was supported in part by DFG (Grant No. KA 4645/1-1).

{\bf A note, added after the paper has been completed.\/}
After our paper was completed, we have learned about \cite{yero_pac}, which covers a broad range of the issues related to the Lamb shift in hydrogen and some other atoms. Concerning the two-loop and three-loop $\alpha^8m$ terms, discussed here, the consideration in \cite{yero_pac} is somewhat different from \cite{codata2014}. In particular, their fit for the two-loop contributions includes $B_{72}$ recently obtained in \cite{our_b72}. The diagrams with vacuum polarization loops in free-loop approximation and the diagrams with closed electron loops beyond the free-loop approximation are considered there separately from the pure self-energy (cf. \cite{yero_num_1}), the same way as we consider them here.

Reference \cite{yero_pac} describes the fit for $G_{60}^{\rm pure\;SE}(Z)$ in few details only. It has a non-physical shape, i.e., comparing with the known shape of the true two-loop function (\ref{G:twoloop}) many logarithmic terms are omitted, and the systematic error is not estimated. The accuracy is worse than the accuracy of the estimation (\ref{eq:b60:j}) for $B_{60}^{\rm pure\;SE}$, which means that the fit was performed rather for the difference $G_{60}^{\rm pure\;SE}\!-\!B_{60}^{\rm pure\;SE}$, than used $B_{60}^{\rm pure\;SE}$ as a free parameter and the result in (\ref{eq:b60:j}) as a data point for $G_{60}^{\rm pure\;SE}(Z=0)$ (as it is done in our paper).
The uncertainty of the diagrams with closed loops beyond the free-loop approximation is based in \cite{yero_pac} on a partial result for $B_{61}^{\rm LbL}$ from \cite{LbL:CS}, while the complete result for $B_{61}^{\rm LbL}$ found here is more than twice larger. As for the three-loop contributions a minor improvement in \cite{yero_pac} was due to use the results on $C_{63}$ and $C_{62}$ from \cite{our_b72}, which are not essential unless the accuracy of constraint on $C_{50}$ is improved first.

\end{document}